\documentclass[pdflatex,sn-mathphys-num]{sn-jnl}% Math and Physical Sciences Numbered Reference Style
\usepackage{graphicx}%
\usepackage{multirow}%
\usepackage{amsmath,amssymb,amsfonts}%
\usepackage{amsthm}%
\usepackage{mathrsfs}%
\usepackage[title]{appendix}%
\usepackage{xcolor}%
\usepackage{textcomp}%
\usepackage{manyfoot}%
\usepackage{algorithm}%
\usepackage{algorithmicx}%
\usepackage{algpseudocode}%
\usepackage{listings}%

\usepackage{lmodern}
\usepackage{xfrac}
\usepackage{subfig}
\usepackage{epsfig}
\usepackage{epstopdf}
\usepackage{textcomp}
\usepackage{setspace}
\usepackage{fullpage}
\usepackage{float}
\usepackage{caption}
\usepackage{siunitx}
\usepackage{tabularx, booktabs}
\usepackage{enumitem}
\usepackage{dcolumn}% Align table columns on decimal point
\usepackage{bm}% bold math
\usepackage[T1]{fontenc}
\usepackage{gensymb}
\usepackage[utf8]{inputenc}
\usepackage{microtype}%
\usepackage{parskip}%

\newcommand{\Jb}{\mathbf{\bar{J}}}
\newcommand{\Jd}{\mathbf{J}}

\theoremstyle{thmstyleone}%
\theoremstyle{thmstyletwo}%

\theoremstyle{thmstylethree}%

\begin{document}

\title[Spinodal decomposition and wetting]{Interplay between spinodal decomposition and wetting in ternary to quinary mixtures}

\author[1]{\fnm{Sandeep} \sur{Kumar}}

\author*[1]{\fnm{Supriyo} \sur{Ghosh}}\email{supriyo.ghosh@mt.iitr.ac.in}

\affil[1]{\orgdiv{Metallurgical and Materials Engineering Department}, \orgname{Indian Institute of Technology Roorkee}, \orgaddress{\city{Roorkee}, \postcode{247667}, \state{Uttarakhand}, \country{India}}}

\abstract{
We use two-dimensional phase-field simulations based on the Cahn--Hilliard description to study spinodal decomposition in $N$-component mixtures with $N$ = 3, 4, 5. In particular, we consider surface-directed spinodal decomposition in which one of the coexisting phases serves as a filler particle, while the remaining components phase separate around its surface. We investigate both nonwetting and wetting conditions by varying the interfacial tensions between the particle and the other components. The competition between wetting kinetics and spinodal decomposition produces rich morphological evolution, including the formation of lattice, droplet, and target (or core-shell) morphologies. The target pattern consists of alternating component-rich rings around the particle. In the intermediate-to-late time regime, these structures undergo further morphological transitions driven by curvature-induced depletion of wetting layers and subsequent coarsening of the surrounding domains, resulting in the breakup of target patterns into isolated domains of the matrix components. We find that domain coarsening slows down systematically as the number of components increases, while wetting accelerates it in all cases. The resulting morphologies and phase-separation pathways demonstrate how particle--component interactions can be used to control microstructural evolution in multicomponent mixtures, with potential implications for systems ranging from industrial materials to cellular biology.
}


\maketitle

\section{Introduction}\label{sec_intro}

The ability to produce materials with a specified morphology is highly desirable since the final mesoscale structure determines many of the properties of the material. One important example is the target or core-shell morphology, which may develop in metallic alloys, polymer blends, and block copolymers~\cite{Karim,clarke2002target}. Such targeted morphologies are often produced via surface-directed spinodal decomposition (SDSD), in which phase separation occurs at or near the surface of ``filler'' additives. These fillers may be present in the matrix in the form of spherical particles, patterned substrates, networks, walls, or objects of arbitrary shape~\cite{Puri_review,ghosh2024_fractal}. They are known to play a critical role in determining the processability, morphology, and properties of the material. Thus, an understanding of the mixture-filler interaction and its ramifications for the resulting equilibrium and nonequilibrium morphologies and phase-separation pathways is of significant practical interest~\cite{balazs_review,Hore,Tanaka2006}.

Spinodal decomposition (SD) of binary mixtures, both in the bulk and under surface-directed conditions, has been studied extensively using experiments, theory, and simulations, and is thus well-known to produce bicontinuous or matrix/dispersed-phase morphologies (e.g., droplets, fibers, or lamellae), depending on the mixture composition~\cite{paul2012polymer,strobl2007polymer}. In contrast, studies of SDSD in multicomponent mixtures are much more limited: those of $N = 3$ systems are relatively limited~\cite{ghosh_pccp,ghosh2020_jcp,ghosh2024_fractal,ghosh2025_dimensionality,shek2022spontaneous}, while those involving $N > 3$ components are rare~\cite{mao2019phase}. In SDSD, particles often interact preferentially with one component, typically the one with lower interfacial energy, leading to wetting phenomena. However, neutral particles can sit directly at the domain interface. The most common finding is that the surface is first enriched with the component (say, $A$, as the core) that has the lower surface free energy. This triggers the formation of an adjacent, concentric layer rich in another component (say, $B$ as the first shell), which in turn can lead to the formation of a third component-rich layer as the second shell, and so on. The growing composition waves around the particles can exhibit constructive and destructive interference, which can affect the stability of these ``target'' patterns. Such core-shell morphologies can significantly alter phase-separation dynamics due to ``frustration'' in interactions arising from the additional components. Wetting-induced target patterns can be useful in templated self-assembly, coatings, and the design of anisotropic materials for optical, sensing, and biomedical applications. Despite this importance, SDSD in multicomponent mixtures remains poorly understood.

Our focus here is therefore to systematically study the effect of a single isolated filler particle on phase separation in $N =$ 3, 4, and 5 component mixtures having symmetric (equiatomic) compositions. For consistency, we treat the particle as a $C$-rich equilibrium phase of the same $N$-component mixture and render it immobile. The remaining components then phase-separate around it, producing the constituent phases. The affinity of each component for the particle depends on the corresponding interfacial energy, which can be readily tailored through a suitable choice of interaction energies and gradient energy coefficients in the phase-field model~\cite{cahn1961spinodal}. This approach allows us to study the interplay between surface-energy-driven phenomena (e.g., wetting) and interfacial-curvature-driven phenomena (e.g., domain coarsening or Ostwald ripening~\cite{voorhees1985}) during spinodal decomposition. We first simulate the phase separation of the corresponding multicomponent mixtures in the absence of a particle, under both nonwetting and wetting conditions, to establish the reference behavior. We then introduce a nonwetting particle for which all components have the same affinity, followed by a wetting system in which the interaction parameters are graded so that the components have a monotonically increasing interfacial energy against the particle. Comparison among these four systems allows us to isolate the effects of the particle and of preferential wetting on the resulting morphologies and phase-separation pathways.

\section{N-component phase-field model}\label{sec_method}
As an extension of our previous works~\cite{ghosh_pccp,ghosh2020_jcp,ghosh2024_fractal}, we consider $N$ components with local volume fraction of component $i$, $c_i$, at position $\mathbf{r}$ and time $t$ subject to $\sum_{i=1}^{N} c_i(\mathbf{r},t) = 1$. The free energy functional follows the Cahn--Hilliard description~\cite{cahn}
\begin{equation}\label{eq_F}
\mathcal{F}
=
N_V\int_V
\left[
f(c_1,\ldots,c_N)
+
\sum_{i=1}^{N}\kappa_i|\nabla c_i|^2
\right]dV ,
\end{equation}
where $\kappa_i$ is the gradient-energy coefficient. The bulk free-energy density is described using the Flory--Huggins regular-solution model~\cite{Porter},
\begin{equation}\label{eq_f}
f=
\frac{1}{2}\sum_{i\neq j}\chi_{ij}c_ic_j
+
\sum_i c_i\ln c_i ,
\end{equation}
where $\chi_{ij}$ represents the pairwise interaction parameter between components $i$ and $j$.
The temporal evolution of each component follows the continuity equation
\begin{equation}\label{eq_continuity}
\frac{\partial c_i}{\partial t}
=
-\nabla\cdot\Jb_i ,
\end{equation}
where $\Jb_i$ is the net mass flux. Following Kramer \emph{et
al.}~\cite{kramer}, the net flux is the diffusion flux plus the amount of $i$
carried by the vacancy flux,
\begin{equation}\label{eq_kramer}
\Jb_i = \Jd_i + c_i \Jd_V, \space
\Jd_V = -\sum_{k=1}^{N} \Jd_k, \space
\Jd_i = -M_i \nabla\mu_i ,
\end{equation}
so that $\sum_i \Jb_i = 0$ by construction. 

Eliminating component $N$ using $\sum_i c_i=1$ gives, for
$i=1,\ldots,N-1$,
\begin{equation}\label{eq_flux_general}
\Jb_i
=
-\bar{M}_{ii}\nabla\tilde{\mu}_i
+
\sum_{j\neq i}^{N-1}
\bar{M}_{ij}\nabla\tilde{\mu}_j ,
\space
\tilde{\mu}_i=\mu_i-\mu_N ,
\end{equation}
where
\begin{align}\label{eq_mobility_general}
\bar{M}_{ii}
&=
(1-c_i)^2M_i
+
c_i^2\sum_{k\neq i}M_k ,
\\
\bar{M}_{ij}
&=
(1-c_i)c_jM_i
+
c_i(1-c_j)M_j
-
c_ic_j\sum_{k\neq i,j}M_k .
\end{align}
The bare Onsager coefficients are taken to be composition dependent,
\begin{equation}\label{eq_bare}
M_i=M_i^0c_i ,
\end{equation}
consistent with $M_i=D_ic_i/RT$. We use $M_i^0=1$ for the mobile components.

The effective chemical potentials obtained from the variational derivative of Eq.~\eqref{eq_F} are
\begin{equation}\label{eq_mu_general}
\tilde{\mu}_i
=
\frac{\partial f}{\partial c_i}
-
2(\kappa_i+\kappa_N)\nabla^2c_i
-
2\kappa_N
\sum_{j\neq i}^{N-1}\nabla^2c_j ,
\end{equation}
with
\begin{equation}\label{eq_dfdc_general}
\frac{\partial f}{\partial c_i}
=
\ln c_i-\ln c_N
+
\sum_{j\neq i}^{N-1}
(\chi_{ij}-\chi_{jN})c_j
+
\chi_{iN}(c_N-c_i).
\end{equation}
The resulting $N-1$ coupled Cahn--Hilliard equations for $i=1,\ldots,N-1$ are
\begin{equation}\label{eq_eom_general}
\frac{\partial c_i}{\partial t}
=
\nabla\cdot
\left[
\bar{M}_{ii}\nabla\tilde{\mu}_i
-
\sum_{j\neq i}^{N-1}
\bar{M}_{ij}\nabla\tilde{\mu}_j
\right] .
\end{equation}

Two-dimensional phase-field simulations are carried out using a semi-implicit Fourier-spectral scheme~\cite{Shen} on a $512\times512$ lattice with periodic boundary conditions in both directions. For consistency, $C$ is chosen as the particle phase in all simulations, since it is the third component in the ternary system. We start with a system containing a $C$-rich fixed particle in the matrix and then allow the surrounding homogeneous matrix to phase separate by solving Eqs.~\eqref{eq_eom_general}. An adaptive time step with a maximum target value between $\Delta t=0.0005$ and $0.05$ is used depending on $\chi_{ij}$. A random noise of $\pm 0.005$ is added to the initial matrix composition to trigger phase separation. We first simulate the corresponding homogeneous systems without a particle, followed by systems containing nonwetting and preferentially wetting particles. The interaction and gradient-energy parameters used for the different systems are summarized in Table~\ref{tab_param}.

To render the particle immobile, the $C$ composition field is constrained to remain at its initial spatial profile, such that $\partial c_C/\partial t=0$. For $N=4$ and $5$, we set $M_C^0=0$ in Eq.~\eqref{eq_bare} and the corresponding row and column of the reduced mobility matrix to zero. Thus, the particle acts as an inert obstacle, while its wetting interaction with the mobile components is retained through the $\chi_{iC}$ and $\kappa_i$ terms. The initial particle composition is taken as the equilibrium composition of the $C$-rich phase obtained from the common-tangent construction for the corresponding $\chi_{ij}$ values, so that it is initially in phase equilibrium with the surrounding matrix and therefore neither dissolves nor coarsens. Finally, to identify the phases in the simulations, we employ $K$-means clustering~\cite{knn_hartigan1979} in the $N$-dimensional composition space to translate local compositions into phase indicator functions. The coarsening kinetics are quantified from the structure function $S_i(k)$ of each component, from which the characteristic domain size $R_i$ is obtained from its first moment~\cite{ghosh2020_jcp,ghosh2025_dimensionality}.
\begin{table*}[t]
\centering
\footnotesize
\setlength{\tabcolsep}{5pt}
\caption{Simulation parameters and the resulting interfacial energies. All simulations are run at the same quench depth $\chi/\chi_c = 1.5$, with $\chi_c$ equal to the number of bulk components. The corresponding interfacial energies ($\sigma_{iC}$) between each bulk phase and the particle phase ($C$-rich) are given. The $\sigma_{AC} = \sigma_{BC}$ in the nonwetting systems suggests that the particle has no selective affinity. In contrast, the wetting case gives $\sigma_{AC} < \sigma_{BC} < \sigma_{CD} < \sigma_{CE}$, and thus $A$ is the wetting component in all simulations.}\label{tab_param}
\begin{tabular}{c c l p{0.26\textwidth} c c c c c}
\toprule
$N$ & particle & condition & $\chi_{ij}$ & $\kappa_i$ &
$\sigma_{AC}$ & $\sigma_{BC}$ & $\sigma_{CD}$ & $\sigma_{CE}$ \\
\midrule
3 & no  & nonwetting & 4.5 (all)                      & 8 & 3.497 & 3.497 & --    & --    \\
  & no  & wetting    & 3.0 (AB) / 3.5 (AC) / 5.5 (BC) & 8 & 2.237 & 3.251 & --    & --    \\
  & yes & nonwetting & 3.0 (AB) / 4.5 (AC, BC)        & 8 & 3.104 & 3.104 & --    & --    \\
  & yes & wetting    & 3.0 (AB) / 3.5 (AC) / 5.5 (BC) & 8 & 2.237 & 3.251 & --    & --    \\
\midrule
4 & no  & nonwetting & 6.0 (all)                      & 8 & 5.120 & 5.120 & 5.120 & --    \\
  & no  & wetting    & 3.5 (AB, AC, BD) / 6.0 (AD, BC) / 8.0 (CD)
                                                      & 8 & 2.307 & 3.719 & 4.944 & --    \\
  & yes & nonwetting & 4.5 (all)                      & 8 & 3.189 & 3.189 & 3.189 & --    \\
  & yes & wetting    & 3.5 (AB, AC, BD) / 6.0 (AD, BC) / 8.0 (CD)
                                                      & 8 & 2.307 & 3.719 & 4.944 & --    \\
\midrule
5 & no  & nonwetting & 7.5 (all)                      & 8 & 6.438 & 6.438 & 6.438 & 6.438 \\
  & no  & wetting    & 4.0 (AB, AC, BD, DE) / 5.0 (AD, BC, BE) / 6.0 (AE, CD) / 7.0 (CE)
                                                      & 8 & 2.790 & 3.408 & 3.988 & 4.658 \\
  & yes & nonwetting & 6.0 (all)                      & 8 & 5.013 & 5.013 & 5.013 & 5.013 \\
  & yes & wetting    & 4.0 (AB, AC, BD, DE) / 5.0 (AD, BC, BE) / 6.0 (AE, CD) / 7.0 (CE)
                                                      & 8 & 2.790 & 3.408 & 3.988 & 4.658 \\
\bottomrule
\end{tabular}
\end{table*}
%------------------------------------------------------------------------------
\section{Results and Discussion}\label{sec_results}
\begin{figure}[htbp]
\centering
\includegraphics[scale=0.5]{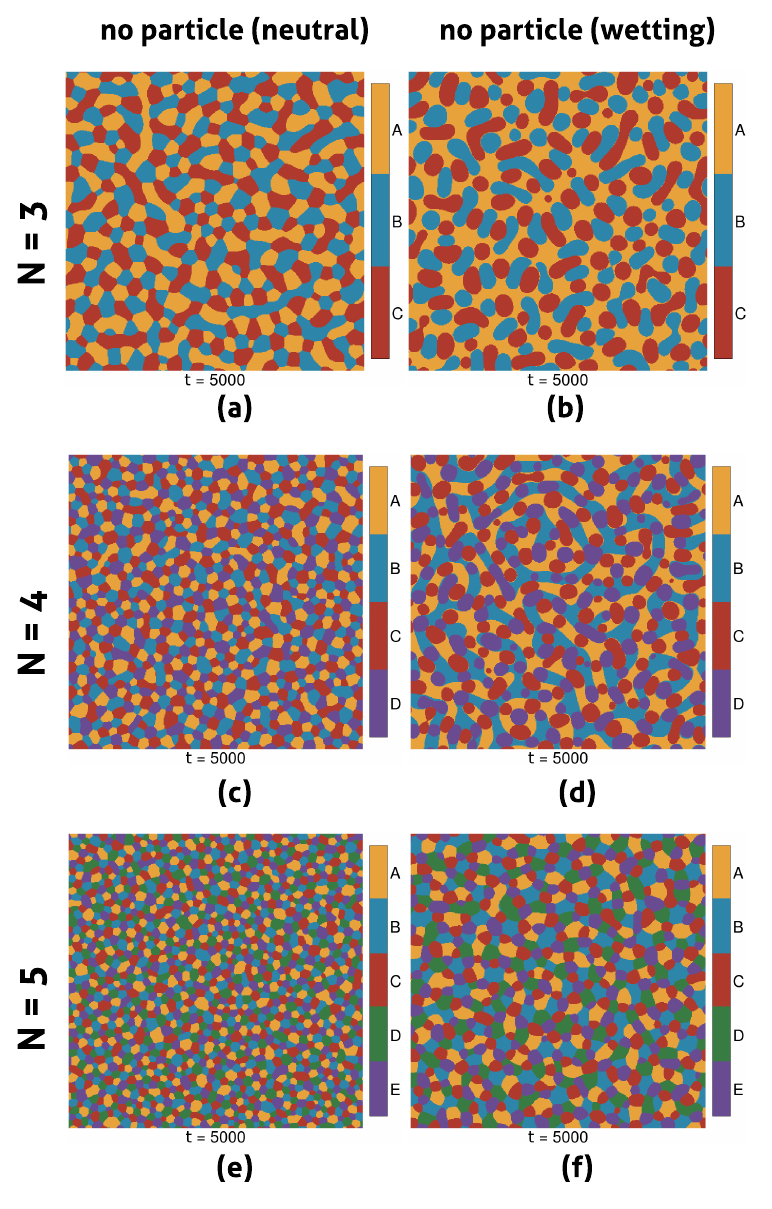}
\caption{Spinodal patterns at simulation time $t$ in ternary ($N = 3$), quaternary ($N = 4$), and quinary ($N = 5$) systems \textit{without a particle}, under nonwetting and wetting conditions (Tab.~\ref{tab_param}).}\label{fig_noparticle}
\end{figure}

\begin{figure}[htbp]
\centering
\includegraphics[scale=0.5]{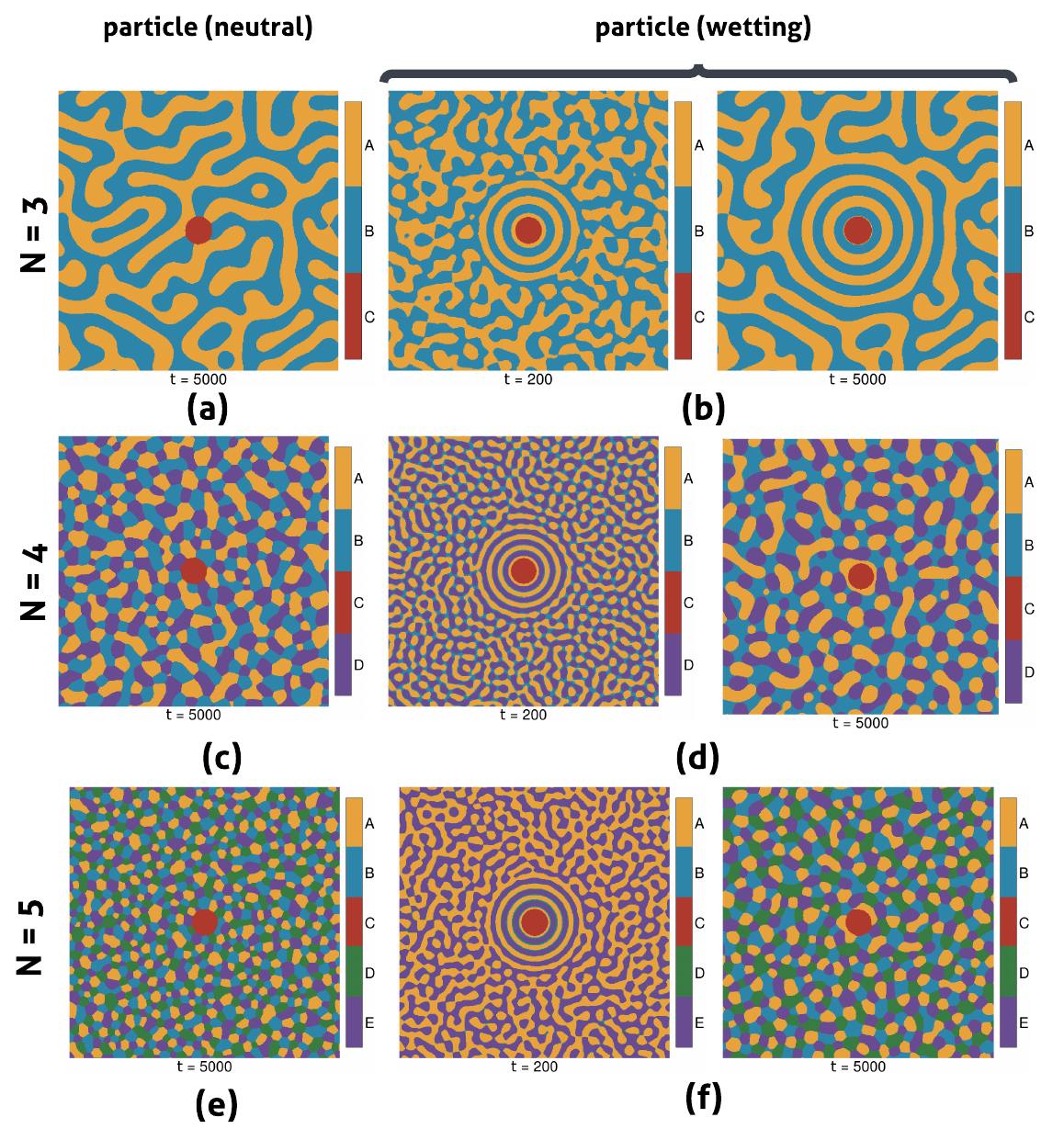}
\caption{Spinodal patterns at simulation time $t$ in ternary ($N = 3$), quaternary ($N = 4$) and quinary ($N = 5$) systems \textit{with an immobile particle}, under nonwetting and wetting conditions (Tab.~\ref{tab_param}). For the wetting systems, morphologies at early ($t = 200$) and intermediate ($t = 5000$) times are shown to illustrate the evolution of the target pattern.}\label{fig_particle}
\end{figure}

\begin{figure}
\centering
\includegraphics[scale = 0.45]{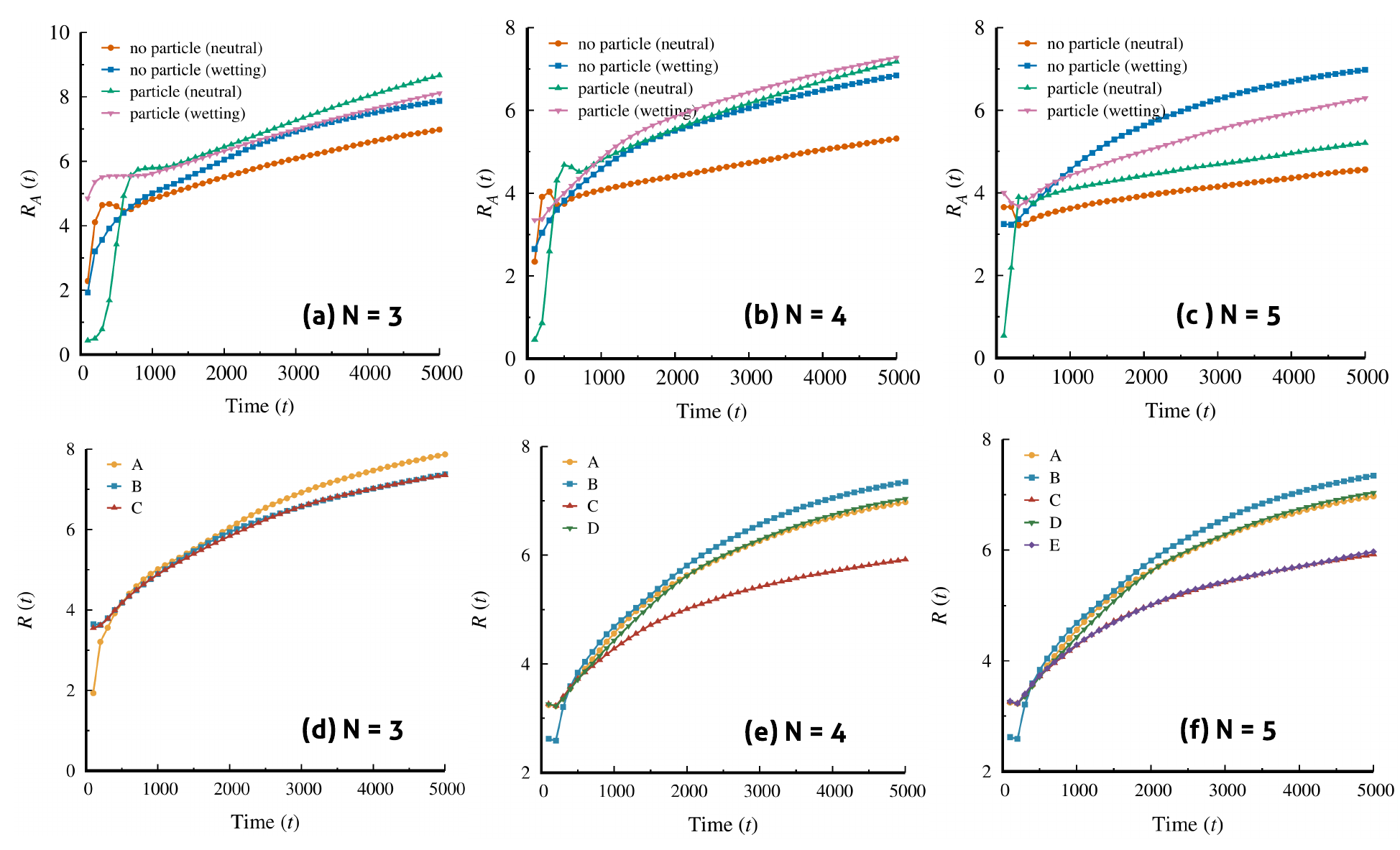}
\caption{Top row: domain size of the wetting component $A$ as a function of time in all systems. Bottom row: domain size of every component in the wetting systems without a particle. The same coarsening trends are observed in the wetting systems with a particle, except for the particle phase (component $C$), which is inert in our study.}\label{fig_domain}
\end{figure}

We work with a compositionally symmetric (equiatomic) alloy mixture and study its decomposition under nonwetting and wetting conditions. First, we perform simulations without particles in a ternary ($N = 3$) mixture (Fig.~\ref{fig_noparticle}a). The system undergoes normal spinodal decomposition (SD), in which all three components begin to phase-separate simultaneously, developing a so-called lattice morphology. Here, the domains corresponding to the three components are interspersed, and multiphase coarsening results from the rearrangement and coalescence of similar domains. The same system under wetting conditions results in the characteristic droplet morphology of the nonwetting phases, associated with higher interfacial energy, in a continuous $A$-rich matrix (Fig.~\ref{fig_noparticle}b). These droplet domains promote efficient Ostwald ripening and become more rounded over time to minimize the interfacial length of the system. Such interfacial coarsening is often mediated by the evaporation--condensation (or Lifshitz--Slyozov--Wagner) mechanism~\cite{lifshitz1961kinetics,Wagner1961}, which is the situation we consider here. The microstructural development becomes significantly different in the presence of a particle (Fig.~\ref{fig_particle}). In nonwetting systems, the particle does not affect bulk phase separation, which becomes binary-like and leads to the well-known bicontinuous morphology; the particle simply resides at an $A$/$B$ interface (Fig.~\ref{fig_particle}a). In wetting systems, the phase separation pathway is completely different (Fig.~\ref{fig_particle}b). This is due to $\sigma_{AC} < \sigma_{BC}$, for which a wetting layer of the preferred component $A$ forms around the particle, followed by a layer of $B$, and so on, forming the target or core-shell morphology; with only two mobile phases, the shell sequence is clear-cut. Phase separation then proceeds into the bulk, which is essentially binary $AB$, leading to the typical bicontinuous morphology. Comparing the target pattern between $t = 200$ and $t = 5000$, the rings become fewer and considerably wider, so they coarsen along with the surrounding matrix. Also, the outermost rings become wavy and progressively merge into the bulk. Over time, the innermost wetting layer also shrinks and may eventually disappear, owing to the curvature-induced depletion associated with the Gibbs--Thomson effect~\cite{Porter}. Wetting often leads to the formation of more rounded droplet domains, thereby reducing interfacial energy and increasing the average interdomain distance (Fig.~\ref{fig_domain}a, d).

In a quaternary ($N = 4$) mixture, phase separation without particles leads to the typical lattice morphology (Fig.~\ref{fig_noparticle}c), with four components interspersed with each other, followed by multiphase coarsening. Wetting promotes continuity of the domains, and hence the phases associated with low interfacial energy (say, $A$ and $B$) form a bicontinuous pattern (Fig.~\ref{fig_noparticle}d). To further minimize the free energy, $C$ and $D$ remain as droplets within the $A$ and $B$ domains, respectively, since the $C$ and $D$ phases are the most incompatible ($\sigma_{CD} = 4.944$, the largest of all pairs), and they progressively become rounded over time. Such phase ordering is due to $\sigma_{AC} = \sigma_{BD} = 2.307$ against $\sigma_{BC} = \sigma_{AD} = 3.719$. Domain continuity is thus most pronounced at $N = 4$, where the tendency of wetting to promote connectivity is not yet outweighed by the tendency of additional components to break the structure into isolated domains. In the presence of a nonwetting particle, the mixture develops a lattice morphology of the bulk phases and hence becomes ternary-like (Fig.~\ref{fig_particle}c). In the wetting system (Fig.~\ref{fig_particle}d), a target pattern with several layers of multiphase shells develops surrounding the particle, with the radial ordering following $\sigma_{AC} < \sigma_{BC} < \sigma_{CD}$. The shells at early time are predominantly made up of $A$ and $D$-rich phases, with $B$ appearing as thin films at the $A$/$D$ boundaries, consistent with $\sigma_{AB}$ and $\sigma_{BD}$ both being smaller than $\sigma_{AD}$. Over time, the number of shells decreases, and they eventually dissolve, owing to curvature-induced coarsening of the circular wetting layers. However, the specific depletion sequence depends on the $\sigma_{ij}$ profile and time. On the other hand, normal SD continues in the bulk, developing a multiphase droplet morphology, followed by domain coarsening. In all systems, the presence of additional components slows the diffusion of species, leading to smaller isolated domains than in the ternary mixture (Fig.~\ref{fig_domain}b, e). Also, wetting leads to more efficient coarsening of domain patterns, resulting in a larger average domain size.

With $N = 5$ mixtures, domain interconnection becomes rare, and phase separation leads to a five-phase isolated lattice or droplet-type morphology, even in the wetting systems (Fig.~\ref{fig_noparticle}e, f). Here, the lower-energy interfaces (such as $AB$, $AC$, $BD$) become more prominent than the high-energy interfaces (such as $CE$, $CD$, $AE$), thereby enhancing subsequent interfacial coarsening. In the presence of a nonwetting particle (Fig.~\ref{fig_particle}e), the phase separation in the matrix essentially becomes quaternary-like and develops a four-phase lattice morphology. In the wetting system (Fig.~\ref{fig_particle}f), multilayer, concentric shells appear surrounding the particle as a core, purely due to $\sigma_{AC} < \sigma_{BC} < \sigma_{CD} < \sigma_{CE}$. The shells are predominantly made up of $A$- and $E$-rich phases, with $B$ and $D$ appearing as thin films at the $A$/$E$ boundaries, consistent with $\sigma_{AB}$, $\sigma_{BD}$ and $\sigma_{DE}$ all being lower than $\sigma_{AE}$. Over time, the preferred component ($A$) progressively becomes depleted from the particle--matrix interface due to curvature effects, and eventually all the rings in the target pattern dissolve and merge into the matrix, and only a partial $A$-rich layer may remain in contact with the particle surface at later times. In the bulk, a multiphase droplet morphology results. In all $N = 5$ mixtures, the domains have a smaller average interdomain distance than in the $N = 3$ and $N = 4$ mixtures, due to sluggish multicomponent diffusion at comparable times. Wetting accelerates phase separation around the particle and the subsequent multiphase coarsening, yielding a larger characteristic domain size than in the nonwetting systems (Fig.~\ref{fig_domain}c). However, the coarsening rates indicated by the slopes of the curves remain similar, and hence the coarsening law in itself is not significantly altered in all systems (Fig.~\ref{fig_domain}f).

We must admit that the present model in the current study is generic and is not tied to a specific statistical model or material. For simplicity, we neglect hydrodynamic effects, which are expected to have only a limited influence at the early and intermediate stages of phase separation reported here. Although we have varied only $\chi_{ij}$ to change the surface energy, variation of $\kappa_i$ can also be used for this purpose (Tab.~\ref{tab_param}). In future work, we will further elaborate on the effects of changes in the above parameters, as well as those of mixture composition and particle concentration, on the stability of core-shell patterns, nested shell structures, and Janus droplets arising from complex wetting conditions (\textit{i.e.}, surface tension inequalities). We also plan to investigate particle mobility and multiparticle effects in these multiphase systems. In the latter case, the growing compositional waves around the particles can exhibit constructive and destructive interference, providing a potentially powerful means of controlling the morphology and stability of the multiphase mixture.

\section{Conclusions}\label{sec_conclusions}
\begin{itemize}
\item We have developed a phase-field model to simulate morphological evolution and subsequent domain coarsening during multicomponent, multiphase spinodal decomposition in $N$ = 3, 4, 5 mixtures under wetting and nonwetting conditions.

\item Wetting promotes continuity of the domains, with incompatible phases embedded within them, whereas increasing $N$ promotes the formation of discrete islands, leading to a lattice morphology in nonwetting systems and droplet morphology in wetting systems.

\item Wetting in particle-containing mixtures leads to a multiphase, multilayer target or core-shell morphology, with the sequence and thickness of the wetting layers depending on the specific interfacial energies between the phases. However, not every component forms a resolved shell; components with low mutual interfacial energy instead segregate to the boundaries between the principal shells.

\item The final morphological transition is driven by the curvature-induced depletion of the wetting layers, and the target pattern eventually breaks up into droplets or elongated domains in the matrix.

\item Wetting accelerates phase separation around particles and subsequent domain coarsening, yielding a larger characteristic domain size. Coarsening becomes progressively slower in multiphase systems as $N$ increases. Collectively, wetting, the particle, and $N$ affect the coarsening kinetics of the mixture in the present study.
\end{itemize}
%---------------------------------------------------------------------------------------- 

\backmatter

%\bmhead{Supplementary information}
%
%If your article has accompanying supplementary file/s please state so here. 
%
%Authors reporting data from electrophoretic gels and blots should supply the full unprocessed scans for key as part of their Supplementary information. This may be requested by the editorial team/s if it is missing.
%
%Please refer to Journal-level guidance for any specific requirements.

%\bmhead{Acknowledgements}

%\section*{Declarations}

%\subsection*{Conflict of interest}
%The authors declare no Conflict of interest.
%\subsection*{Data availability}
%The data that supports the findings of the study are available from the corresponding author upon reasonable request.

\bibliography{spinodal}% common bib file
%% if required, the content of .bbl file can be included here once bbl is generated
%%\input sn-article.bbl

\end{document}